\documentstyle[12pt,aaspp4]{article}

\begin{document}
 
\title{The Discovery of Two Nearby Carbon Dwarfs}
\author{Patrick J. Lowrance\altaffilmark{1}, 
J. Davy Kirkpatrick\altaffilmark{1}, I. Neill Reid\altaffilmark{2},
Kelle L. Cruz\altaffilmark{3}, \& James Liebert\altaffilmark{4}}
\altaffiltext{1}{Infrared Processing and Analysis Center, California Institute of Technology, Pasadena, CA; lowrance,davy@ipac.caltech.edu }
\altaffiltext{2}{Space Telescope Science Institute, Balitmore, MD; inr@stsci.edu}
\altaffiltext{3}{University of Pennsylvania, Philadelphia, PA; kelle@hep.upenn.edu}
\altaffiltext{4}{University of Arizona, Tucson, AZ; liebert@as.arizona.edu }

\begin{abstract}

The comparison of optical and 2MASS near-infrared photometry for large
samples of catalogued proper motion stars has the potential to discover
previously unrecognized nearby objects of rare type. In this paper we present 
the discovery of two new carbon dwarfs, LSR 2105+2514 and LP 758-43,
which were drawn from proper motion lists and which lie in a sparsely
populated part of optical/near-IR color-color space. Their optical
spectra, exhibiting absorptions by C$_2$ and/or CN, are discussed. 
LSR 2105+2514 is believed to lie within 200 pc and would have
$M_{K_s} \ge 6.7$, making it lower in luminosity than any carbon
dwarf with a measured trigonometric parallax. LP 758-43, which is
not as red but still probably cooler than the best studied carbon
dwarfs, is believed to lie within 360 pc. Using our optical/near-infrared 
selection technique on published lists of proper motion stars, we hope 
in the near future to expand further the current sample of carbon dwarfs, 
which numbers only 31 objects at this writing.

\end{abstract}

%\keywords{Stars: atmospheres --- stars:carbon --- techniques: photometric}

%\keywords{Stars: atmospheres --- stars:carbon --- techniques: photometric}

\section{Introduction}

The distinctive optical spectra of carbon giants have led to their use as 
probes of the Galactic halo, especially since their bright magnitudes enable
radial velocities to be obtained at very large distances (Aaronson 1983).  
For the first three quarters of the last century, giants were the only
luminosity class of carbon stars known. With the discovery of 
G77-61 (Dahn et al. 1977) at a mere 58 pc and therefore relatively low 
luminosity, a category of dwarf carbon (dC) stars was finally recognized.
This, however, presented a puzzle because a low mass star is incapable of the 
helium fusion needed to create carbon. This puzzle was resolved
by invoking the presence of an evolved, second member of the system --- now
an undetected white dwarf secondary. It was assumed that the second 
member, as it passed through its
asymptotic giant branch (AGB) phase, transferred mass onto a
lower mass main sequence star, enriching it in carbon-bearing material (Dahn 
et al 1977). The lower
mass member was proposed to be a metal-poor subdwarf, and therefore the
small amounts of oxygen it does contain in the atmosphere will be
overwhelmed by the carbon. A decade after this theory was 
advanced, its credibility was strengthened by the discovery of radial 
velocity variations in G77-61 (indicating an unseen companion; Dearborn 
et al 1986), and by discoveries of 
two other dC stars in binaires containing a visual white dwarf secondary 
(Heber et al 1993; Liebert et al 1994). A review of dwarf carbon star research 
has been given recently by Green (2000).

The process of making a carbon dwarf may also begin to explain the existence
of a rare class of carbon giants known as CH stars. CH stars are halo members 
with peculiar abundances, C/O $>$ 1, and a strong overabundance of s-process 
elements thought to be produced during an AGB phase. Interestingly, they are 
all in known binary systems, and those with derived mass ratios 
are consistent with 
white dwarf secondaries. As such, these giant stars may represent the dC's
at a more advanced evolutionary state (McClure 1984; Wallerstein \& 
Knapp 1998).

Current understanding of dC's is, however, limited by the small number of 
such objects now recognized. Only as more dC's are found can we fully test 
the current hypotheses and understand the role of dC's in the overall picture of 
stellar evolution. Fortunately, large areal photometric surveys are
beginning to uncover these objects in larger numbers. Margon et al (2002) 
have reported the discovery of 39 faint high-latitude carbon stars from 
the Sloan Digital Sky Survey (SDSS), of which 17 exhibit proper motions 
exceeding their three-sigma astrometric uncertainties and are assumed to be
dC's. In this paper we report the first two dC's uncovered by the Two Micron All Sky
Survey (2MASS).

\section{The First New Carbon Dwarf}

\subsection{Discovery and Spectroscopic Confirmation}

From an analysis of multi-epoch Digitized Sky Survey (DSS) images lying 
within 25 degrees of the Galactic plane, L{\'{e}}pine, Shara \& Rich (2002; 
hereafter LSR02) published a list of objects having proper motions larger 
than 0.5$\arcsec$/yr and magnitudes down to R=19.8 mag. 
Their list, although a rediscovery of many objects tabulated in the 
Luyten Half-Second Catalogue (Luyten 1979), 
also included 141 new proper motion stars not recognized previously.
Approximately forty percent of those objects fall in the 2MASS Second 
Incremental Data Release (Cutri et al 2000), so their characteristics
can be studied via diagrams employing the B and R data from the DSS as well
as the JHK$_s$ data from 2MASS. In looking at the colors of these objects
we noticed that one, LSR2105+2514 
(21:05:16:58,+25:14:48.1; J2000), has a very red J$-$K$_s$ color (1.28) like
a late-M or early-L dwarf (Table 1) but an R$-$J color (1.6) that is far
too blue for such a dwarf. A finder chart for this object is shown in LSR02.

We observed LSR 2105+2514 spectroscopically 
on 07 Aug 2002 UT using the Double Spectrograph (Oke \& Gunn 1982) on the 
Hale 5-meter telescope at Palomar Observatory. Our instrumental setup 
used the D68 dichroic to split the light between the two channels at
$\sim$6800 \AA. A 300 lines/mm grating was used in the blue arm and a 316 
lines/mm grating in the red arm with grating tilts that provided a few hundred
Angstroms of duplicate coverage between the two arms. This provided 
continuous wavelength coverage from 3800 to 9100 \AA. Use of 
a 2{\farcs}0 slit
resulted in a resolution of 8 \AA. Observations were reduced using standard 
techniques following Kirkpatrick, Henry, \& McCarthy (1991) and flux 
calibrated using the standard star Feige 110 (Hamuy et al 1994). 

The resulting spectrum is shown in Figure 1. The strong bands of C$_2$ at 
4382, 4737, 5165, 5636, and 6191 \AA\ are easily recognizable, indicating
that this is a carbon star. The high proper motion measured by LSR02 of 
0.563$\arcsec$/yr along with its magnitudes of B=18.7$\pm$0.5 and
R=16.1$\pm$0.5 indicate that LSR 2105+2514 is a dwarf star and not a 
background giant.

\subsection{Known Carbon Dwarfs and a Comparison to LSR 2105+2514}

In a given volume of space, carbon dwarfs are believed to greatly
outnumber carbon giants and as such should be 
the dominant type of carbon star in
the Galaxy (Green 2000). Nonetheless, few dC's are
catalogued. Before the 17 dC's recently announced by Margon et al
(2002), previous discoveries only dribbled in since the 1970's.
Table 1 gives a list of all known dwarf carbon stars known as of this
writing. 

Only three of the carbon dwarfs in Table 1 have known distances, and the 
faint magnitudes and presumably larger distances of many of 
the others may preclude a 
robust trigonmetric parallax determination. 
LSR 2105+2514 is bright enough that it has already been added to the USNO 
parallax program. Even without a $\pi_{Trig}$, however, we can place a 
crude upper limit to its distance based on its measured proper 
motion of 0$\farcs$563/yr and $\theta = 150.2 \deg$ (LSR02). 
Assuming that LSR 2105+2514 is gravitationally bound in the Galaxy, it 
cannot exceed an escape velocity of $\sim$ 500 km/s 
(Carney, Latham, \& Laird 1988). We then calculate expected (U,V,W) motions, 
stepping out in distance assuming V$_{rad}=0$. We have transformed 
those velocities to
galactic motions using a solar motion of (9, 11, 6) and assuming a
rotational velocity of 220 km/s for the Local Standard of Rest.
LSR2105+2514 has a total velocity exceeding 500 km/s if the distance
exceeds 200 parsecs. That suggests that LSR2105+2514 has M$_{Ks} \geq$
6.7 mag. This would indicate a lower luminosity carbon dwarf than 
any of the ones with a measured 
$\pi_{Trig}$ (see Table 1). This suggestion is seemingly supported by its red 
J$-$K$_s$ color of 1.28, which indicates it is also one of the coolest carbon dwarfs.

\section{Finding Other Carbon Dwarfs}

Currently, the only way to distinguish between a carbon dwarf and a carbon 
giant is through its luminosity, and hence one needs a measure of 
the parallax or an indirect distance indicator such as proper motion. To date,
discriminators based on spectroscopy or photometry have been proposed, but
none has met with 100\% success. 

To define a list of carbon dwarf candidates, we have chosen the list of
36,085 NLTT objects (Luyten 1980) that Gould \& Salim (2002) have
cross-referenced against the 2MASS 2nd Incremental Data Release. If we
plot known carbon dwarfs on an R$-$J vs. J$-$K$_s$ diagram along with
objects from the NLTT, we find that carbon dwarfs generally have much
redder J$-$K$_s$ colors than the cloud of main sequence stars at similar
R$-$J colors (Figure 2). Examining the infrared spectra of carbon dwarfs
in Joyce (1998), we find shallow CN bands at R and K along with the 
K-band CO absorptions that are normally in low-mass stars, but a defining
characteristic of these carbon dwarfs are large CN absorption bands at J.
This results in bluer R$-$J and redder J$-$K$_s$ colors than main-sequence
objects of similar temperature. Exploring objects from the NLTT in this 
part of color space should increase the sample size of known carbon dwarfs.

Objects with infrared colors are often placed on (J$-$H)/(H$-$K$_s$) to 
decipher between the giant and dwarf populations. Unfortunately, as 
Margon et al 2002 comments and Figure 3 demonstrates, 
some of the carbon dwarfs have J$-$H colors like giants, and therefore, 
cannnot be distinguished using this diagram alone. 
Likewise, there are a few unusual giants that fall in the lower half of the 
diagram along with the bulk of the dwarfs. It is therefore important 
to search for {\it spectroscopic} indicators that might 
help us understand if these differences are due to temperature or abundance effects 
and possibly differentiate carbon dwarfs from carbon giants.

\section{A Second New Carbon Dwarf}

Using the R$-$J vs. J$-$K$_s$ selection criteria outlined above and in the caption to Figure 2, 
we have identified and observed several NLTT carbon dwarf candidates. 
LP 758-43, with R$-$J$=$2.0 and J$-$K$_s$=1.10,    
was observed on 2002 Sep 26 UT with the
Multi-Aperture Red Spectrograph (MARS) on the KPNO Mayall 4m
telescope. We used the VB8050-450 grating and 2.0$\arcsec$-wide
long slit, covering the wavelength range 5600 to 10000 \AA\ at a resolution
of $\sim$9\AA. The data were reduced and wavelength calibrated using
standard techniques in IRAF, and flux calibrated through observations
of the spectrophotometric standard Feige 110. The spectrum of LP 758-43 is 
plotted in Figure 1 and compared with other known carbon dwarfs. The CN bands
appear very similar in depth to the previously known carbon dwarfs. A finding 
chart for LP 758-43 is presented in Figure 4. Other candidates which 
turned out to be close doubles, mismatches with reddened stars, 
a white dwarf/red dwarf binary and M dwarfs will be summarized in the 
complete survey (Lowrance et al 2003, in prep).

With K$_s$=11.90, LP 758-43 is one of the brighter examples of a dC (see Table 1), 
enabling future trigonometric parallax measurements.  
Without a $\pi_{Trig}$, we can attempt to place a crude upper limit to the distance 
using the arguments used for LSR 2105+2514. For its measured proper motion 
of 0.255$\arcsec$/yr and $\theta=90.4$ not to exceed the escape velocity of $\sim$500 km/s, 
its distance would be no greater than $\sim$360 pc with a M$_{K_{s}} > $ 4.1 mag. 
The J-K$_s$ color of LP 758-43 (J-K$_s$=1.1) is 
redder than the average J-K$_s$ color (J-K$_s$=0.95) of the three carbon dwarfs 
with known distances. Those three have an average M$_{K_s} =$ 6.5 mag, so if we assume LP 758-43 has 
a similar or fainter absolute magnitude at K$_s$, it is more likely 
located at less than 120 pc.

\section{Conclusions}

We present the optical spectra of two newly discovered carbon dwarfs, LSR
2105+2514 and LP 758-43, with proper motions of 0.563$\arcsec$/yr and
0.255$\arcsec$/yr, respectively. Both objects were selected based on their 
unusual optical to infrared colors and lie in a locus on the optical/
near-infrared color-color diagram populated by other known carbon
dwarfs. Looking through proper motion catalogs - where it is assured 
that faint objects are nearby dwarfs and not background giants - for 
objects of similar color will allow us to produce lists of carbon
dwarf candidates for spectroscopic follow-up. Finding more carbon dwarfs, 
of which only 31 are currently known, can help us understand the 
role of dC's in stellar evolution.  Finally, broadening the
sample will allow a search for spectroscopic discriminants that could help
distinguish between carbon dwarfs and carbon giants.

P.J.L. acknowledges support from a National Research Council Fellowship. 
P.J.L. and  J. D. K. acknowledge the support of
the Jet Propulsion Laboratory, California Institute of Technology, which is operated under
contract with the National Aeronautics and Space Administration. 
KLC acknowledges support from a NSF Graudate Student Fellowship. We are grateful to 
the referee H. Harris for a careful review of the manuscript and suggested improvements. 
P.J.L. acknowledges great support from the Palomar Observatory staff. This publication makes use
of data from the Two Micron All-Sky Survey, which is a joint project of the University of
Massachusetts and the Infrared Processing and Analysis Center, funded by the National
Aeronautics and Space Administration and the National Science Foundation. 
%This research
%has made use of the SIMBAD database, operated at CDS, Strasbourg, France.
%This research has made extensive use of the NASA/ IPAC Infrared Science Archive, 
%which is operated by the Jet Propulsion Laboratory, California Institute of Technology, 
%under contract with the National Aeronautics and Space Administration 

\clearpage

\begin{deluxetable}{cccccccc}
\scriptsize
\tablecaption{List of All Known Carbon Dwarfs} \label{cdwarfs}
\tablenum{1}
\tablehead{\colhead{Name} & \colhead{RA (J2000)}  &  \colhead{Dec (J2000)} & \colhead{J} & \colhead{H} & \colhead{K$_{s}$} & \colhead{J-K$_{s}$} & \colhead{Ref\tablenotemark{a}}}
\startdata
LHS 1075 &00 26 00.2 & $-$19 18 52 &  12.54 $\pm$ 0.02 & 11.92 $\pm$ 0.03 & 11.59 $\pm$ 0.03 & 0.95 & 2 \\ 
SDSS J0039373$+$152911&00 39 37.3 & $+$15 29 11 & ... & ... & ... &  & 7 \\
WIE93 0041$-$295&00 43 34.9 & $-$29 18 08 & ...   &  ...  &  ...   &  & 10  \\
WIE93 0045$-$259 &00 48 17.6 & $-$25 38 38 &  ...  &  ...  &  ...   &  &10 \\
SDSS J0121503$+$011303&01 21 50.3 & $+$01 13 03 & 15.11 $\pm$ 0.05 & 14.25 $\pm$ 0.05 & 13.81 $\pm$ 0.05 & 1.30 & 7 \\
SDSS J0125267$+$000449 &01 25 26.7 & $+$00 04 49 & ...   &  ...  & ...    &  &7 \\  
%SDSS J0130071$+$002635 &01 30 07.1 & $-$00 26 35 & 15.59 $\pm$ 0.07 & 14.89 $\pm$ 0.08 & 14.35 $\pm$ 0.08 & 1.24 & 7 \\
SDSS J0256346$-$084854 &02 56 34.6 & $-$08 48 54  &  ...  &   ... &   ...  &  &7 \\  
G77$-$61, LHS 1555 &03 32 38.0 & $+$01 58 00 & 11.47$\pm$  0.02 & 10.84 $\pm$  0.02 & 10.48 $\pm$  0.02 & 0.99 & 1 \\
SDSS J0736213$+$390725 &07 36 21.3 & $+$39 07 25 & 16.76$\pm$ 0.14  & 16.09  $\pm$ 0.21 & 15.44 $\pm$ 0.15 & 1.33 &7 \\  
SDSS J0822514$+$461232 &08 22 51.4 & $+$46 12 32 & 15.50 $\pm$0.06 & 14.68 $\pm$0.07 & 14.48 $\pm$ 0.09   & 1.02 &7 \\
SDSS J0826268$+$470912 &08 26 26.8 & $+$47 09 12 &   15.78 $\pm$0.07 & 15.19 $\pm$ 0.08 & 15.02 $\pm$0.11  & 0.76 & 7 \\
PG 0824+289B\tablenotemark{b} &08 27 05.1 & $+$28 44 02 &  ... & ... & ... & & 4 \\
SDSS J0858533$+$012243 &08 58 53.3 & $+$01 22 43 & 15.82  $\pm$ 0.10 & 15.11  $\pm$ 0.09  & 14.39 $\pm$0.08 & 1.43 &  7 \\
SDSS J0900114$-$003606 &09 00 11.4 & $-$00 36 06 & ...   & ...   &  ...   &  & 7 \\  
SDSS J0948587$+$583020 &09 48 58.7 & $+$58 30 20  & ...   &  ...  & ...    &  &7 \\ 
SDSS J1004325$+$004338 &10 04 32.5 & $+$00 43 38  &  ...  & ...   &  ...   &  &7 \\ 
CLS 29 &10 40 06.4 & $+$35 48 02 & 12.98 $\pm$ 0.03 & 12.29 $\pm$ 0.03 & 12.00 $\pm$ 0.02 & 0.98  & 9 \\ 
CLS 31 &10 54 29.6 & $+$34 02 30 &  15.50 $\pm$ 0.05 & 15.02 $\pm$ 0.07 &  14.66 $\pm$  0.08 & 0.84 & 2  \\
KA$-$2 &11 19 03.9 & $-$16 44 50 &  13.24 $\pm$ 0.02 & 12.61$\pm$ 0.03 &  12.48 $\pm$ 0.03  & 0.76 & 8 \\
SDSS J1129504$+$003345 &11 29 50.4 & $+$00 33 45& 16.47 $\pm$ 0.16 & 15.69 $\pm$ 0.14 & 15.52 $\pm$ 0.24 & 0.95 & 7 \\
SDSS J1147317$+$003724 &11 47 31.7 & $+$00 37 24  & ...   &  ...  &   ...  & & 7 \\ 
CLS 50 &12 20 00.8 & $+$36 48 03 & 14.41 $\pm$ 0.03 & 13.93 $\pm$  0.03 & 13.79 $\pm$ 0.04  & 0.62 & 3  \\
SDSS J1353330$-$004039 &13 53 33.0 & $-$00 40 39& 14.61 $\pm$ 0.04 & 13.80 $\pm$ 0.04 & 13.61 $\pm$ 0.05 & 1.00 & 7 \\
SDSS J1421124$-$004823 &14 21 12.4 & $-$00 48 23  & ...   &   ... &   ...  & & 7 \\ 
CBS 311 &15 19 05.9 & $+$50 07 03 & 15.56 $\pm$ 0.06 & 14.75$\pm$ 0.07 & 14.16$\pm$ 0.07 & 1.40 & 6  \\
SDSS J1537322$+$004343 &15 37 32.2 & $+$00 43 43 & 15.20 $\pm$ 0.05 & $\geq$14.36  &   $\geq$14.08 &  $\geq$1.12 & 7 \\
CLS 96, LP 328-57 &15 52 37.5 & $+$29 28 02 & 13.80 $\pm$ 0.03 & 13.20$\pm$ 0.04 & 12.88 $\pm$ 0.03 & 0.92  & 2 \\
WIE93 2048$-$348&20 52 02.6 & $-$34 37 32 & 16.47 $\pm$ 0.14 & 15.87 $\pm$  0.19 &  15.29 $\pm$  0.16 & 1.18 &10  \\
LSR2105+2514 & 21 05 16.6 & +25 14 48 & 14.48 $\pm$ 0.03 &  13.77 $\pm$ 0.03 & 13.20 $\pm$ 0.04 & 1.28  & 5 \\
LP 758-43 &21 49 37.8 & $-$11 38 28 & 13.00  $\pm$ 0.02 & 12.26  $\pm$ 0.02  & 11.90  $\pm$ 0.03 & 1.10 &  5 \\
SDSS J2302550$+$005904 &23 02 55.0 & $+$00 59 04 & 15.61 $\pm$ 0.06 & 14.80 $\pm$ 0.06 & 14.60 $\pm$ 0.08 & 1.01  & 7 \\
\enddata
\tablenotetext{a}{Discovery reference:(1)Dahn et al 1977; (2)Green et al 1994; (3) Green et al 1992; (4) Heber et al 1993; 
(5) this paper; (6) Liebert et al. 1994; (7) Margon et al 2002; 
(8) Ratnatunga 1983; (9) Totten, Irwin, \& Whitelock 2000; (10) Warren et al. 1992;}
\tablenotetext{b}{Photometry does not split into two sources in the 2MASS All-Sky Release. Combined photometry is 
J=12.423$\pm$0.032; H$=$11.802$\pm$0.035; K$_s$=11.650$\pm$0.03} 
\tablenotetext{c}{Photometry is from the 2MASS All-Sky Release. Those without photometry were too faint for 2MASS.}
\tablenotetext{d}{The carbon dwarfs with determined parallaxes (Harris et al 1998) are G77-61 
$\pi_{Trig}$=16.9$\pm$2.2 mas (M$_{K_s}=6.62$), LHS1075 $\pi_{Trig}$=7.96$\pm$0.84 mas (M$_{K_s}=6.09$), and CLS 96 
$\pi_{Trig}$=4.54$\pm$0.66 mas (M$_{K_s}=6.17$)}
\end{deluxetable}

\clearpage

\epsscale{0.8}
\plotone{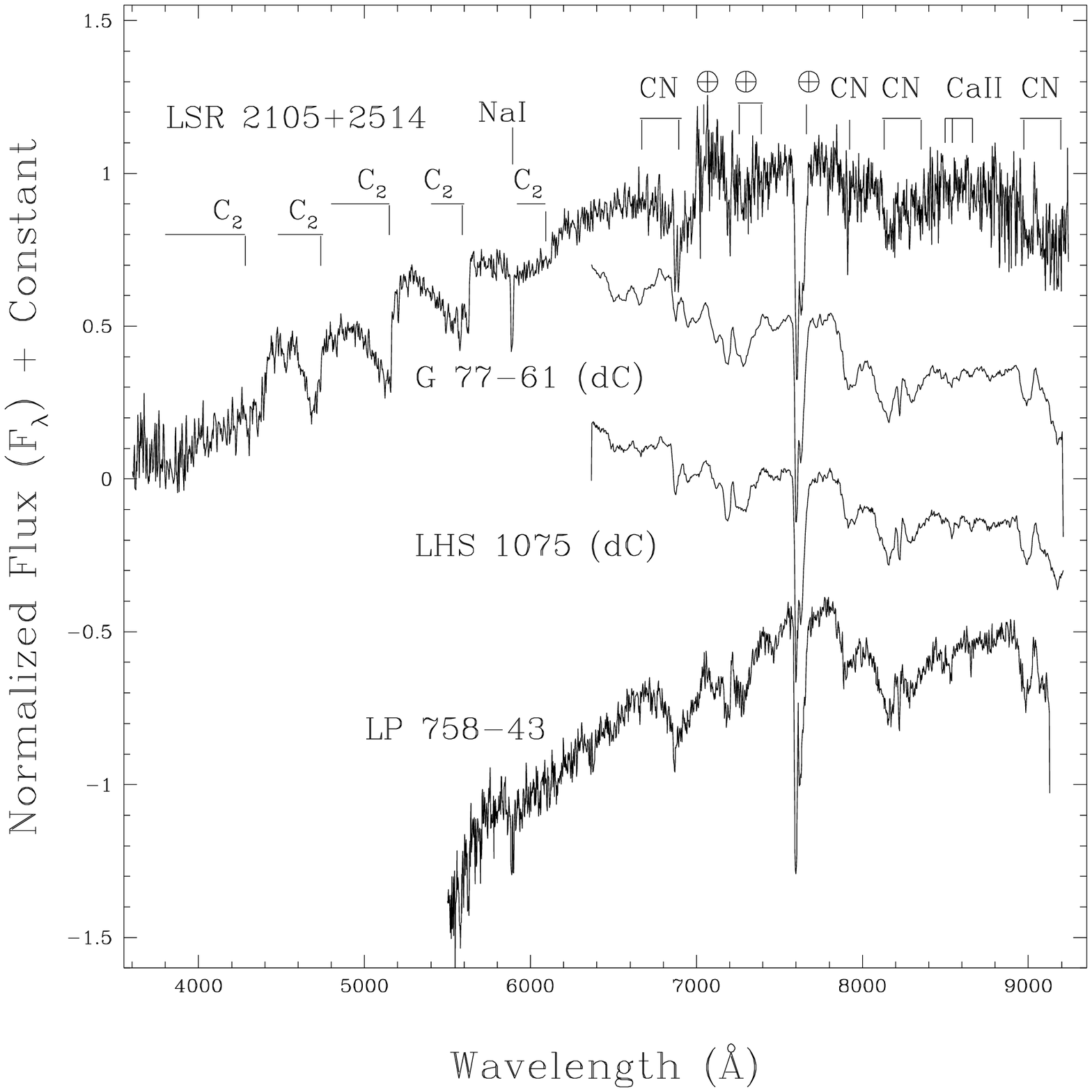}
\figcaption{Optical spectra of LSR2105+2514 (top) and LP 758-43 (bottom) compared with 
known carbon dwarfs G77-61 and LHS 1075 from Kirkpatrick (1992). 
The Swan bands of C$_2$ at 4382,4737,5165, and 5636 \AA\ 
are labelled as well as the sharp bandhead of C$_2$ at 6191\AA\ not often seen in
carbon giants. The Na I ``D'' doublet, Ca II triplet, and the CN bands are also 
labelled. In the LSR2105+2514 spectrum, 
there was a small calibration problem 
in the blueward side of the red channel ($\sim$6800$-$7200\AA) that caused 
the flux in that region to be incorrectly elevated. This small portion is 
suspect, but the remainder of the spectrum should not be affected. The spectra are all 
normalized at 7500\AA, and offsets of -0.5, -1.0, and -1.5 have been applied 
to the normalized flux levels of G77-61, LHS 1075, and LP 758-43, respectively,
to separate them from each other in the figure.}

\plotone{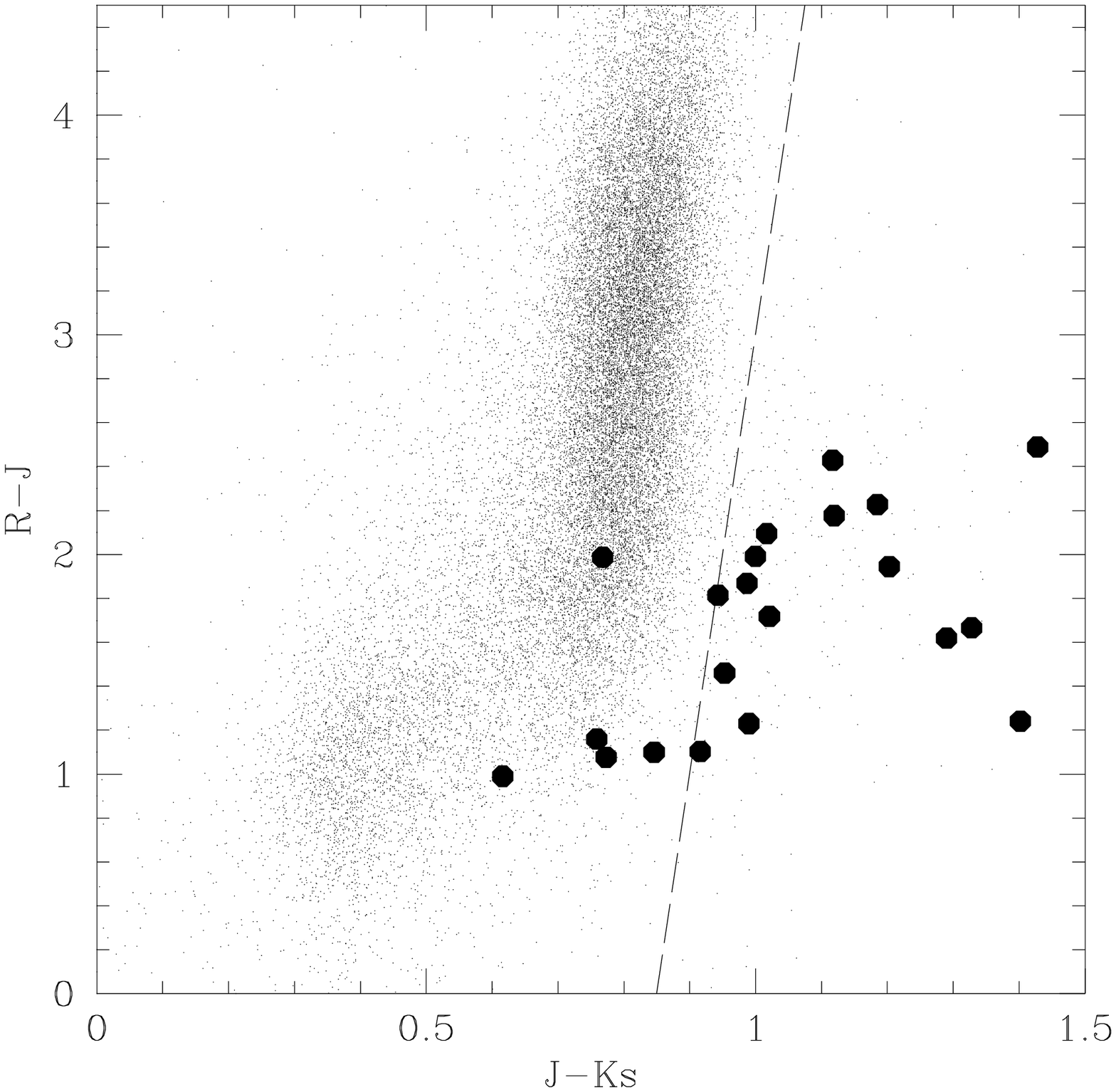}
\figcaption{R-J vs J-K$_s$ for stars from Gould \& Salim (2002) with the known 
carbon dwarfs (filled circles). The dotted line represents a boundary of 
(J$-$K)$=$ 20 (R$-$J) $-$ 17, used to define a cutoff for the 
observation selection of possible carbon dwarfs.}

\plotone{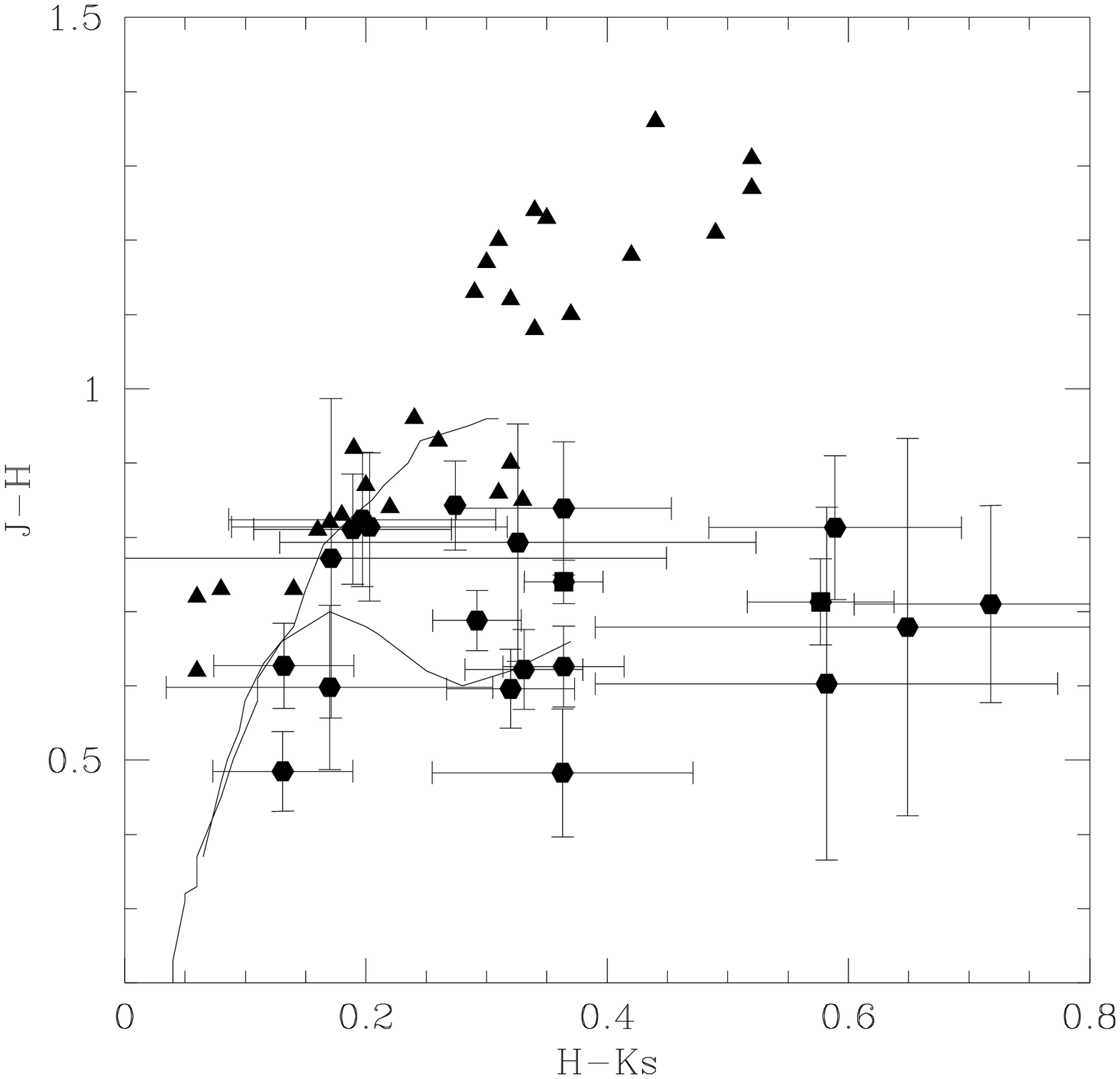}
\figcaption{J$-$H vs H$-$K$_s$ for the two discoveries in this paper (filled squares) and all 
2MASS photometered carbon dwarfs (filled hexagons; the two with error bars greater than 0.2 have been omitted here) 
listed in Table 1 compared with carbon giants (filled triangles). 
Overplotted are the Bessell \& Brett (1988) tracks for giants (top) and dwarfs (bottom). 
Notice some of the carbon dwarfs fall within the giant branch of 
these Bessell-Brett tracks, leading to confusion for objects with small or no proper motion.}

\plotone{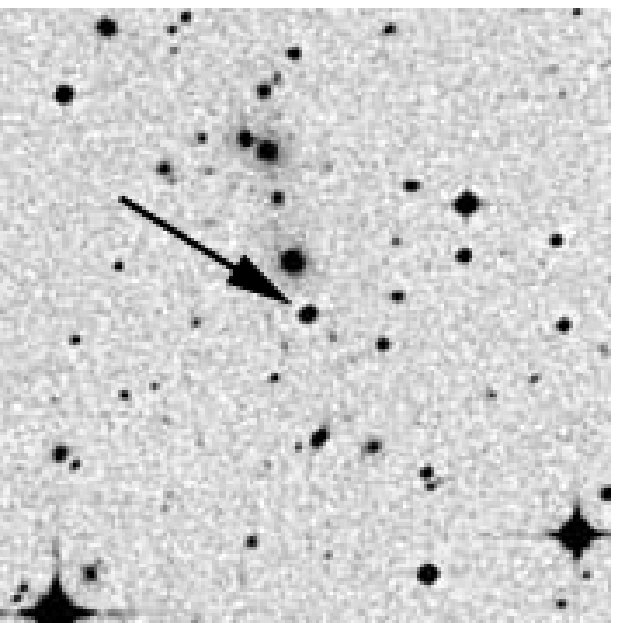}
\figcaption {Finder chart for LP 758-43. The 5x5 arcminute R-band
image from the XDSS (epoch 09 Oct 1988) where 
an arrow distinguishes the carbon dwarf.}

\end{document}